\documentclass[aps,prl,twocolumn,groupedaddress,showpacs]{revtex4}

\usepackage{bm}
\usepackage{graphicx}
\usepackage{color}

\newcommand{\be}{\begin{equation}}
\newcommand{\ee}{\end{equation}}
\newcommand{\bea}{\begin{eqnarray}}
\newcommand{\eea}{\end{eqnarray}}
\newcommand{\lsim}{\raise.35ex\hbox{$<$}\kern-0.75em\lower.5ex\hbox{$\sim$}}
\newcommand{\gsim}{\raise.35ex\hbox{$>$}\kern-0.75em\lower.5ex\hbox{$\sim$}}

\def\mb#1{\mbox{\boldmath $#1$}}

\begin{document}
%
\title{
Saturated Ferromagnetism from Statistical Transmutation in Two Dimensions
}

\author{
Yasuhiro Saiga$^1$ and Masaki Oshikawa$^2$
}

\affiliation{
$^1$Department of Physics, Nagoya University, Nagoya 464-8602, Japan \\
$^2$Department of Physics, Tokyo Institute of Technology, Oh-okayama, Meguro-ku, Tokyo 152-8551, Japan
}

\begin{abstract}
The total spin of the ground state is calculated
in the $U \to \infty$ Hubbard model
with uniform magnetic flux perpendicular to a square lattice,
in the absence of Zeeman coupling.
It is found that the
saturated ferromagnetism emerges
in a rather wide region
in the space of the flux density $\phi$ and the electron density $n_{\rm e}$.
In particular, the saturated ferromagnetism at $\phi=n_{\rm e}$
is induced by the formation of a spin-1/2 boson,
which is a composite of an electron and the unit flux quantum.
\end{abstract}




\maketitle


Ferromagnetism remains a challenging problem in spite of
being among the best known phenomena in condensed matter physics.
In particular, the origin of ferromagnetism in electron systems
is fundamentally quantum-mechanical and non-perturbative~\cite{TasakiReview}.
There are rather few established mechanisms of ferromagnetism,
especially those of the saturated (complete) ferromagnetism.
Saturated ferromagnetism is defined as
the ground state of the many-electron system
{\em with spin-independent interaction} having the maximum possible
total spin.

Nagaoka's theorem is one of few rigorous results on
the saturated ferromagnetism~\cite{Nagaoka}.
It guarantees that
the saturated-ferromagnetic state is the unique ground state
when a single hole is inserted in the half-filled Hubbard model
with infinite on-site repulsion $U$.
Unfortunately, this theorem is limited to the single-hole case.
Numerical studies suggest that Nagaoka ferromagnetism
is unstable in the thermodynamic limit
at finite hole densities~\cite{Putikka-etal}.

The flat-band ferromagnetism is another rigorous
result~\cite{Mielke,Tasaki92}.
Namely, the saturated ferromagnetism is proved rigorously under certain
conditions, in electron systems with a (nearly) flat dispersion
in the lowest band for a single electron.
The ferromagnetism in a system with the low electron density
and singular density of states
near the Fermi level~\cite{Hlubina-etal}
may also be related to the flat-band mechanism.
However, as these results still have limited applicability,
it is worth pursuing other mechanisms of (saturated) ferromagnetism.

The difficulty in realizing the ferromagnetism in many-electron
systems may be attributed to the Pauli principle for electrons.
In the absence of interaction, the ground state of the system
is generally paramagnetic, because the lower energy bands are filled
with up and down spins.
If we consider a system of bosons rather than fermions, the
intrinsic tendency to favor paramagnetism may be absent.
In fact, it was proved that
in a continuous system with spinful bosons,
one of the ground states is always fully polarized
if explicit spin-dependent interactions are absent~\cite{Suto,Eisenberg-Lieb}.
This statement holds also
in a lattice model~\cite{Eisenberg-Lieb,Fledderjohann-etal}.
In particular,
for the infinite-$U$ Hubbard model with spin-1/2 bosons,
the total spin of the ground state is shown to be maximal
{\it for all hole densities},
unlike 
in the electronic Hubbard model~\cite{Fledderjohann-etal}.

Thus, the saturated ferromagnetism could emerge in an electron system
if the statistics of the electron is transmuted to bosonic.
In fact, the statistical transmutation is indeed possible
in two-dimensional (2D) systems~\cite{Semenoff,Fradkin},
and it has been applied to fractional quantum Hall
effect~\cite{Girvin-MacDonald}.

Combining these ideas, we can expect that 
the 2D electron system
in the presence of
an external gauge (magnetic) field
exhibits the (saturated) ferromagnetism
thanks to the formation of the composite boson.
Namely, when the applied magnetic field amounts to unit flux quantum per
electron, the magnetic-flux quantum may be assigned to an electron.
The composite particle consisting of an electron and the attached flux
is then expected to have spin $1/2$
and to obey the Bose statistics with hard-core constraint.
In this way, in the mean-field level, the original system can be mapped
into a spin-1/2 boson system without magnetic field,
which exhibits
the saturated ferromagnetism~\cite{Suto,Eisenberg-Lieb,Fledderjohann-etal}.
However, as the ``flux attachment'' argument
is not rigorous, whether this mechanism actually leads
to the ferromagnetism in an electron system has to be checked.

In quantum Hall systems in the continuum,
the fully spin-polarized ground state is favored
for the filling factor $\nu=1$
(and in general, for $\nu=1/m$ with $m$ odd),
without the Zeeman energy~\cite{Zhang-Chakraborty,Rezayi,Maksym}.
This is referred to as quantum Hall ferromagnets~\cite{MacDonald-etal}.
While this ferromagnetism is usually associated with
the antisymmetric nature of the orbital part of the wavefunction,
it could also be regarded as a
consequence of the formation of a spin-1/2 boson which
is composed of an electron and $m$ flux quanta~\cite{MacDonald-etal}.
On the other hand, the dispersion in quantum Hall systems
in the continuous space is completely flat (Landau levels).
Thus the saturated ferromagnetism may also be understood as
a special case of the flat-band ferromagnetism~\cite{comment2}.
It is not clear whether the statistical transmutation
is essential to realize the saturated ferromagnetism
in quantum Hall systems.

In order to clarify this issue, in this Letter we study
an electron system on a lattice.
We demonstrate an example of saturated ferromagnetism
which is due entirely to statistical transmutation and is distinct
from the flat-band variety.

Let us introduce the $U \to \infty$ Hubbard model
on a square lattice, with the gauge (magnetic) flux $\phi$
per plaquette~\cite{Schofield-etal}:
\bea
{\cal H} &=& - \sum_{\langle ij \rangle \sigma}
\left[ t_{ij} (\phi_{ij}) c_{i \sigma}^{\dagger} c_{j \sigma}
+ {\rm H.c.} \right]
 + U \sum_i n_{i \uparrow} n_{i \downarrow},
\label{HubbardHamiltonian} \\
  && t_{ij} (\phi_{ij}) \equiv
t \exp ({\rm i} 2 \pi \phi_{ij} / \phi_0),\nonumber \\
  && \phi = \sum_{\rm oriented \; plaquette} \phi_{ij},
\ \ 
\phi_0 \equiv h/e \equiv 1, \nonumber
\eea
where $\langle ij \rangle$ refers to the nearest-neighbor pairs.
Periodic boundary conditions are imposed in both directions,
unless explicitly mentioned otherwise.
In the $U \to \infty$ limit, we have
\be
  {\cal H} = - \sum_{\langle ij \rangle \sigma}
\left[ t_{ij} (\phi_{ij}) \tilde{c}_{i \sigma}^{\dagger} \tilde{c}_{j
\sigma} + {\rm H.c.} \right],\label{Hamiltonian}
\ee
where $\tilde{c}_{i \sigma} = c_{i \sigma} ( 1 - n_{i,-\sigma} )$,
which means that double occupancy at each site is excluded.
We stress that, in our model~(\ref{Hamiltonian}),
we have not included the Zeeman coupling of spins to the magnetic field.
The model is therefore completely isotropic in the spin space,
and the total spin is a conserved quantum number.

Exact numerical diagonalization
for $4 \times 4, \sqrt{18} \times \sqrt{18}$, 
and $\sqrt{20} \times \sqrt{20}$ clusters
is employed in our study.
We study the system with various
values of the electron density $n_{\rm e}$ and
the flux per plaquette $\phi$.
In particular, we need to investigate the case  $\phi=n_{\rm e}$
where the statistical transmutation to boson would occur.
Under periodic boundary conditions, the total flux
of the system is quantized to an integer.
Thus $\phi$ can take only integral multiples of $1/N$, where
$N$ is the number of sites (plaquettes).
In order to study all the possible values of $\phi$,
we need to use the string gauge~\cite{Hatsugai-etal}:
Choosing a plaquette $S$ as a starting one,
we draw $N-1$ outgoing arrows (strings) from the plaquette $S$,
so that each plaquette other than $S$ is the endpoint of a string.
Then we set $\phi_{ij}$ on a link $ij$ to $\phi {\cal N}_{ij}$
taking account of the orientation,
where ${\cal N}_{ij}$ is the number of strings cutting the link $ij$.
We have checked that
the single-electron spectrum for a small system size ($N \sim 20$)
in the string gauge
approximately reproduces the Hofstadter butterfly
in the thermodynamic limit~\cite{Hofstadter}.

\begin{figure*}
\includegraphics[width=5.5cm]{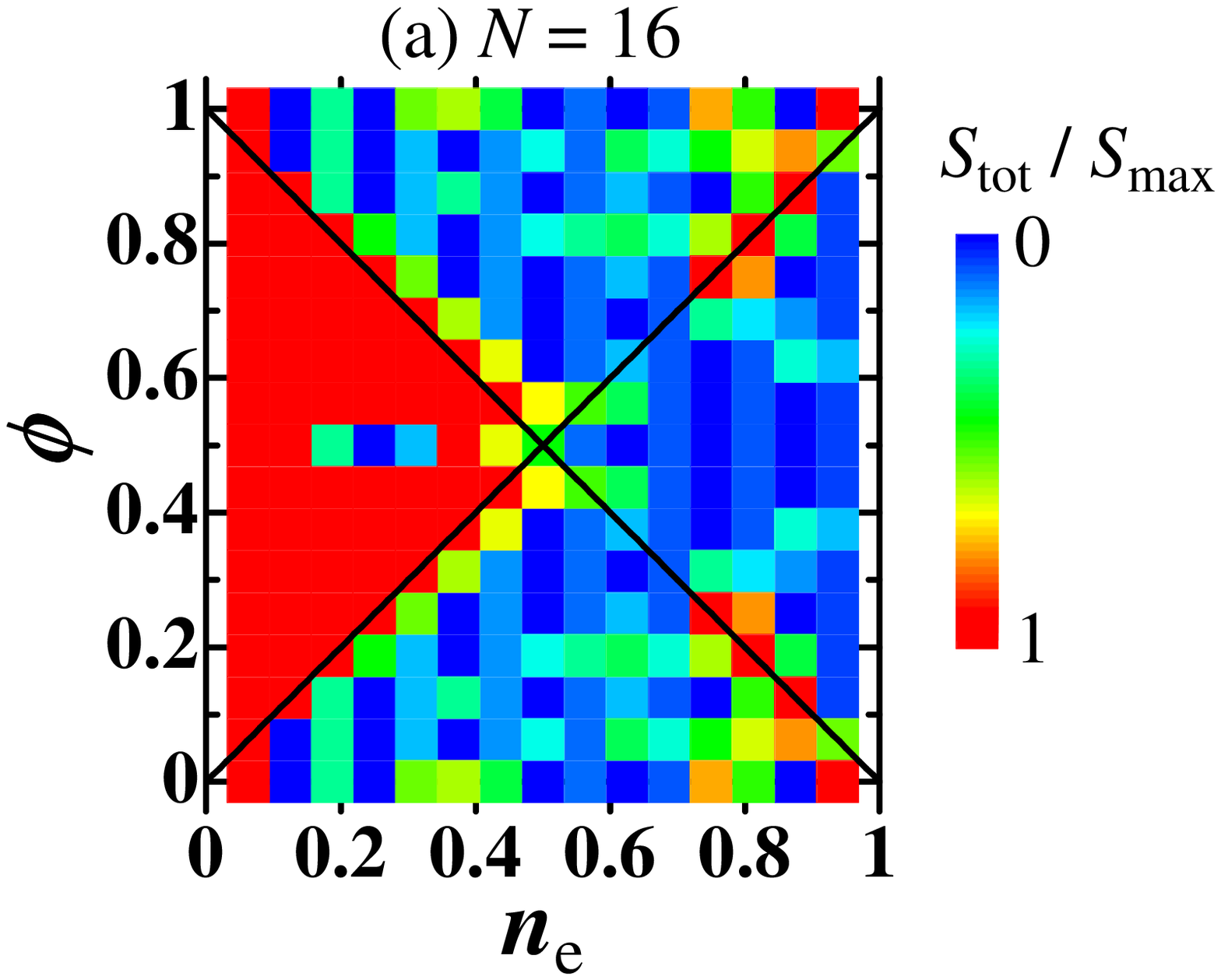}%
\includegraphics[width=5.5cm]{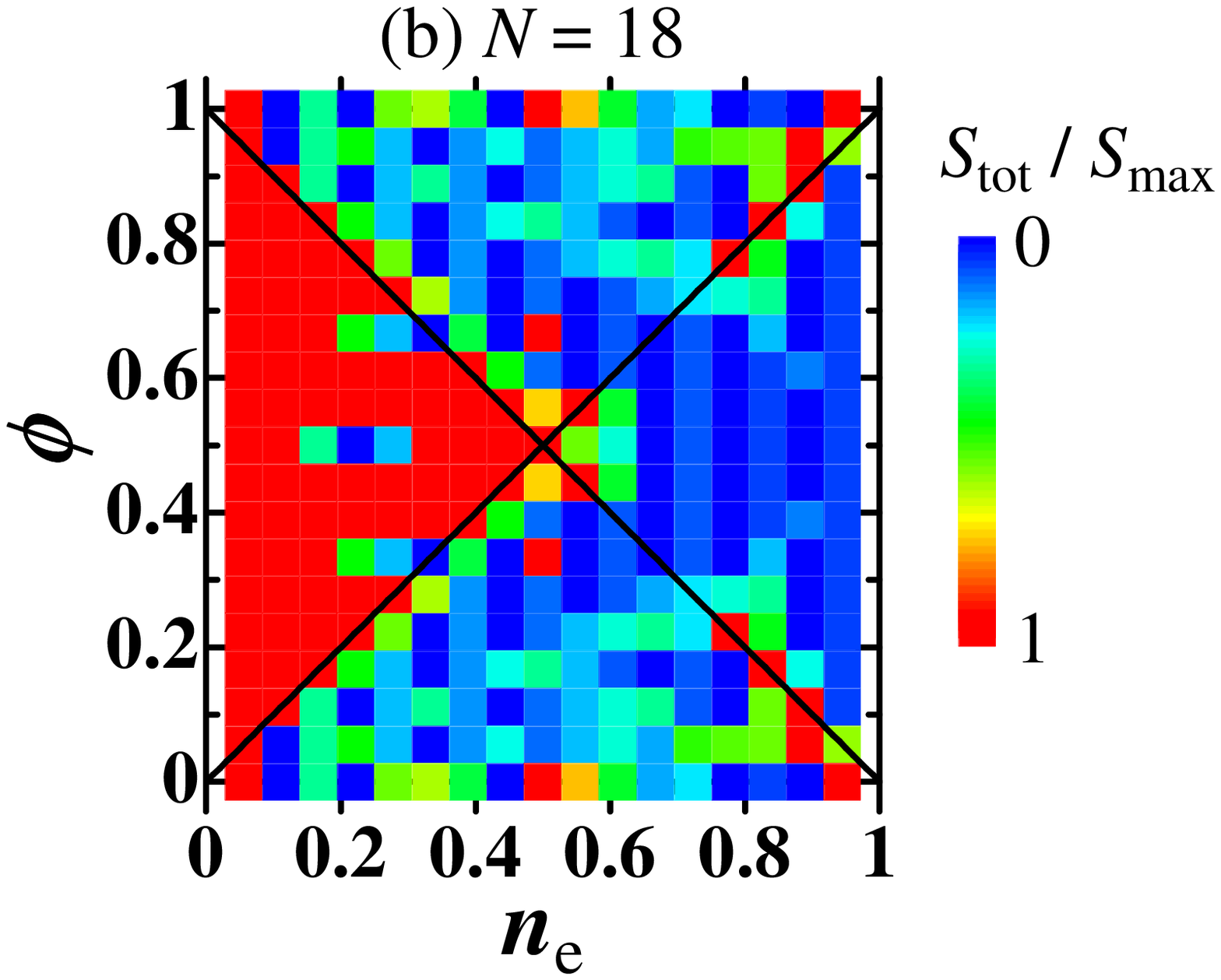}%
\includegraphics[width=5.5cm]{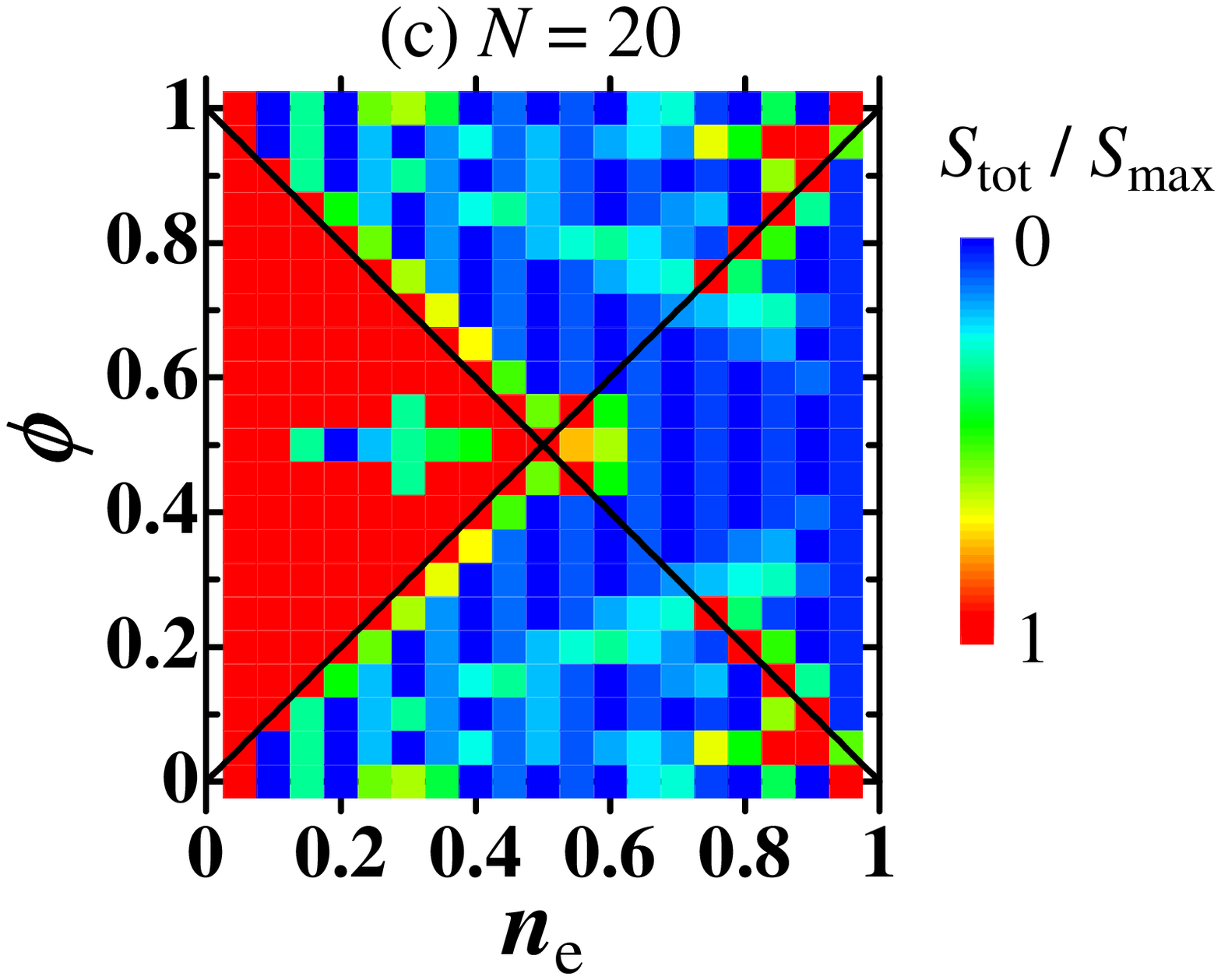}%
\caption{Total spin of the ground state as functions of
the electron density ($n_{\rm e}$) and the flux per plaquette ($\phi$)
in the 2D $U \to \infty$ Hubbard model.
Solid lines are straight lines with $\phi=n_{\rm e}$ and $\phi=1-n_{\rm e}$,
where statistical transmutation is expected.
}
\label{totalspin}
\end{figure*}

The total spin $S_{\rm tot}$ at zero temperature
can be evaluated from
the expectation value of
$( \mb{S}_{\rm tot} )^2 = ( \sum_\ell \mb{S}_\ell )^2$ in the ground state,
where $\mb{S}_\ell$ is the spin operator at site $\ell$.
In Fig.\ \ref{totalspin} we show 
the scaled total spin $S_{\rm tot}/S_{\rm max}$
in the $\phi$--$n_{\rm e}$ plane.
Here $S_{\rm max}$ is given by $N_{\rm e}/2$
with $N_{\rm e}$ being the number of electrons.
Red regions correspond to saturated-ferromagnetic states.
In addition to Nagaoka ferromagnetism in the single-hole case with $\phi=0$,
we find two common features irrespective of the system size.
(i) Saturated ferromagnetism appears along a straight line
with $\phi = n_{\rm e}$ (or $\phi = 1 - n_{\rm e}$)
except $0.6 \; \lsim \; n_{\rm e} \; \lsim \; 0.7$.
(ii) Saturated ferromagnetism appears in the triangular region
surrounded by three straight lines:
$n_{\rm e} = 0$, $\phi = n_{\rm e}$, and $\phi = 1 - n_{\rm e}$.

The result (i) confirms the expectation based
on the statistical transmutation.
Moreover, we find that, in most cases of $\phi = n_{\rm e}$,
the saturated ferromagnetism
is robust against twisting the boundary condition.
This is in contrast to Nagaoka ferromagnetism,
where the total spin of the ground state is changed from
maximum to zero as the boundary is twisted~\cite{Kusakabe-Aoki}.
Besides, we have checked for $N=16$ and $18$ that
the saturated-ferromagnetic ground state at $\phi = n_{\rm e}$
is nondegenerate except for the trivial
$2 S_{\rm max} + 1$-fold degeneracy.
This is consistent with the ferromagnetism for
spin-1/2 bosons~\cite{Eisenberg-Lieb,Fledderjohann-etal}.

In order to distinguish the ferromagnetism due to the statistical
transmutation from possible ``flat-band'' varieties,
we define the spectral functions
\bea
  && D^- (\omega) = \frac{1}{N} \sum_{\ell,n} | \langle \Psi_n (N_\uparrow-1, N_\downarrow; \phi) | c_{\ell \uparrow} | \Psi_0 (N_\uparrow, N_\downarrow; \phi) \rangle |^2 \nonumber \\
  && \times \delta [ \omega + E_n (N_\uparrow-1, N_\downarrow; \phi) - E_0 (N_\uparrow, N_\downarrow; \phi) + \mu ],\label{Dminus} \\
  && D^+ (\omega) = \frac{1}{N} \sum_{\ell,n} | \langle \Psi_n (N_\uparrow+1, N_\downarrow; \phi) | c^\dagger_{\ell \uparrow} | \Psi_0 (N_\uparrow, N_\downarrow; \phi) \rangle |^2 \nonumber \\
  && \times \delta [ \omega - E_n (N_\uparrow+1, N_\downarrow; \phi) + E_0 (N_\uparrow, N_\downarrow; \phi) + \mu ].\label{Dplus}
\eea
Here $\mu$ is the chemical potential,
and $| \Psi_n (N_\uparrow, N_\downarrow; \phi) \rangle$
denotes an eigenstate with energy $E_n (N_\uparrow, N_\downarrow; \phi)$
in the system with $N_\uparrow$ up-spins,
$N_\downarrow$ down-spins, and the flux $\phi$.
We define the index $n$ so that $n=0$ corresponds to
the ground state with the given $N_\uparrow$ and $N_\downarrow$.
In the following, we set $N_\uparrow = N_\downarrow = N_{\rm e}/2$
for $N_{\rm e}$ even.
$D^{\pm}$ can be estimated numerically
by the continued-fraction method~\cite{Gagliano-Balseiro}.
Below, we show the results for two values of $n_{\rm e}$;
$n_{\rm e}=4/20$ and $n_{\rm e} = 18/20$ as representatives
of the ``low electron density'' and ``high electron density'' regimes,
respectively.

\begin{figure}
\includegraphics[width=5cm]{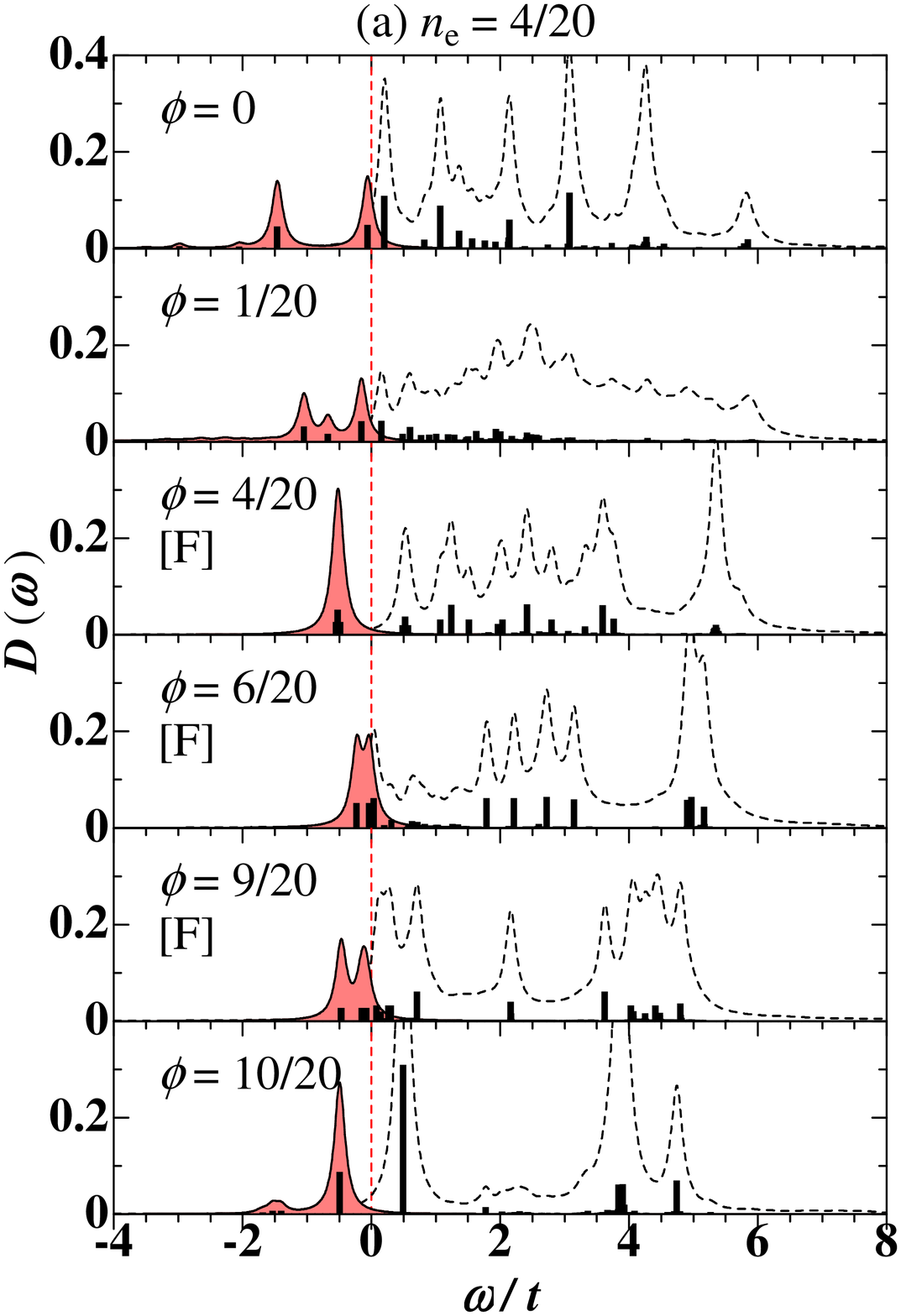}

\vspace{0.2cm}

\includegraphics[width=5cm]{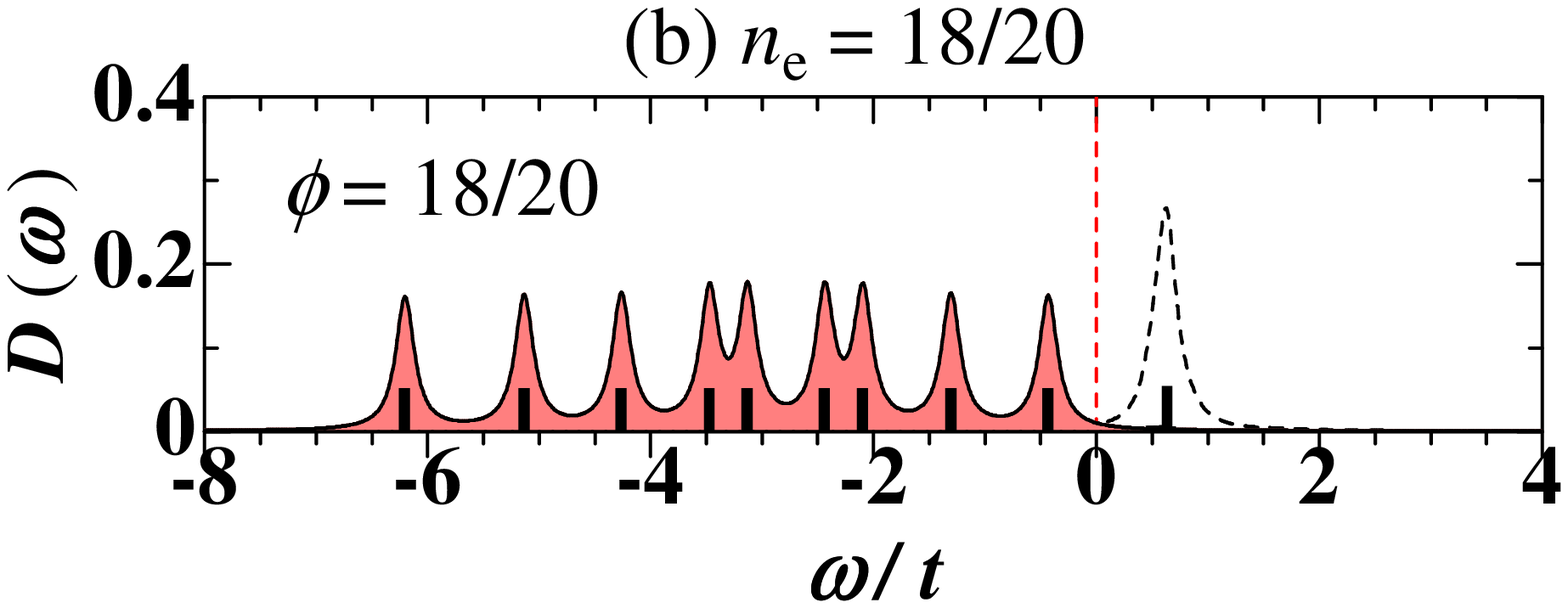}

\caption{Spectral functions $D^{\pm}$
in the 2D $U \to \infty$ Hubbard model.
(a) $n_{\rm e}=4/20$ with various values of $\phi$;
(b) $n_{\rm e}=18/20$ with $\phi=18/20$.
The delta functions (vertical bars) are broadened by a Lorentzian with
a width of $0.1t$.
The chemical potential is located at the zero energy.
Cases with [F] exhibit saturated ferromagnetism.
Colored regions represent $D^-$,
while the dashed curves show $D^+$.}
\label{DOS}
\end{figure}

First let us focus on the ``low electron density'' case,  $n_{\rm e}=4/20$.
Figure \ref{DOS}(a) shows the evolution of $D^{\pm}(\omega)$
with varying $\phi$.
The sum of $D^+$ and $D^-$ corresponds to the density of
states (either occupied or unoccupied).
Apparently it is always spread over a similar
range of energy, representing the ``bandwidth'' which
is about $8t$ although there is some $\phi$-dependence.

On the other hand, at this density,
we find a crucial difference in the spectral
function $D^-$ corresponding to magnetism.
Namely, $D^-$ is concentrated in a narrow range of energy when
the system exhibits the saturated ferromagnetism
for $\phi=4/20,5/20,\cdots,9/20$ (and $1 - \phi$).
In contrast, when the saturated ferromagnetism is absent, 
$D^-$ is spread over a region of energy.
Intuitively, the spectral function $D^-$
corresponds to the density of states {\em occupied by electrons}.
The narrow distribution of $D^-$ compared to the ``bandwidth''
indicates a variant of the narrow or nearly flat-band
ferromagnetism~\cite{Hlubina-etal}.

In fact, there is more difference in $D^-$ between the
cases with and without saturated ferromagnetism,
than what is visible in Fig.~\ref{DOS}.
$D^-$ vanishes completely below a certain threshold
($\omega/t = -0.550, -0.229, -0.474$ respectively for
$\phi=4/20, 6/20$ and $9/20$) when the system exhibits
the saturated ferromagnetism.
In contrast, there is a continuous ``shoulder'' of
low intensity (invisible in Fig.~\ref{DOS})
down to much lower energy $\omega/t \sim - 20$
when saturated ferromagnetism is absent.
This seems to be consistent again with our interpretation.

In particular,
for $\phi=4/20 (= n_{\rm e})$, the spectral function $D^-$
is localized within a narrow energy band below an apparent gap
around the Fermi level.
This appears similar to the quantum Hall ferromagnet
at $\nu=1$ in the continuum~\cite{MacDonald-etal}.
In this case, the statistical transmutation mechanism and
the flat-band mechanism appear indistinguishable.

Now let us discuss the saturated ferromagnetism observed in the
``high electron density'' regime, at $\phi = n_{\rm e} =18/20$.
Figure \ref{DOS}(b) shows the spectral functions in this case.
Clearly, the spectral function $D^-$ spreads over
almost the entire bandwidth
even though the system does exhibit
the saturated ferromagnetism.
Thus the ferromagnetism at $\phi = n_{\rm e} =18/20$
is difficult to be understood in terms of the flat-band mechanism,
and appears to be exclusively due to the statistical transmutation
mechanism.
In fact, changing the value of $\phi$ destroys the saturated
ferromagnetism at this electron density, as expected from the
statistical transmutation scenario.

\begin{figure}
\includegraphics[width=7cm]{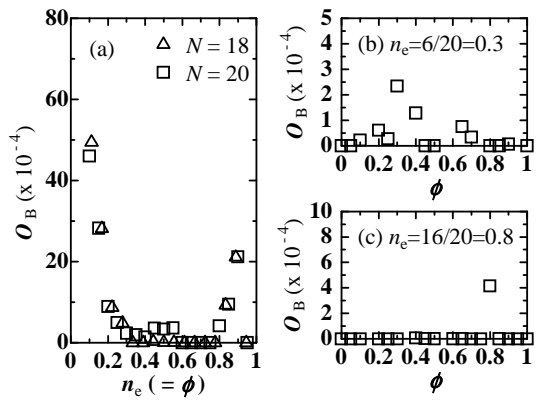}%
\caption{Order parameter of spin-1/2 bosons in the 2D $U \to \infty$ Hubbard model.
(a) $n_{\rm e}(= \phi)$-dependence; (b) $\phi$-dependence for $n_{\rm e}=6/20$;
(c) $\phi$-dependence for $n_{\rm e}=16/20$.
In (b) and (c), data are not plotted where the initial and/or final states in Eq.\ (\ref{OB}) have degeneracy.}
\label{orderparameter}
\end{figure}

In order to further confirm the statistical transmutation scenario
at $\phi = n_{\rm e}$, 
we define the following operators:
$b_{\ell \sigma} = {\rm e}^{- {\rm i} {\cal J}_\ell} c_{\ell \sigma}$
and
$b^\dagger_{\ell \sigma} = c^\dagger_{\ell \sigma} {\rm e}^{{\rm i} {\cal J}_\ell}$,
where
${\cal J}_\ell = - m \sum_{i (\ne \ell)} \theta_{\ell i} n_i$
with
$n_i = \sum_{\sigma = \uparrow,\downarrow} c^\dagger_{i \sigma} c_{i \sigma}$.
In general, $m$ denotes the number of magnetic-flux quanta,
and we set $m=1$ in the present case.
$\theta_{\ell i}$ is the argument of the vector drawn from site $i$
to site $\ell$.
Note that the relation $\theta_{\ell i} - \theta_{i \ell} = \pm \pi$ holds
for $i \ne \ell$.
Then we can prove that these operators satisfy the commutation relations
of bosons for two different sites.
There is a large freedom in determining explicit values of $\theta_{i \ell}$.
Here we follow a prescription introduced in Ref.\ \cite{Cabra-Rossini},
although this prescription unavoidably breaks translation invariance
of a periodic cluster.
The order parameter for the condensation of the composite bosons
may be defined by
\be
  O_{\rm B} = \frac{1}{N} \sum_\ell | \langle \Psi_0 (N_\uparrow-1,N_\downarrow; \phi - \frac{1}{N}) | b_{\ell \uparrow} | \Psi_0 (N_\uparrow,N_\downarrow; \phi) \rangle |^2.\label{OB}
\ee
We again set $N_\uparrow = N_\downarrow = N_{\rm e}/2$
for $N_{\rm e}$ even,
and $N_\uparrow = N_\downarrow + 1 = (N_{\rm e}+1)/2$
for $N_{\rm e}$ odd.
Figure \ref{orderparameter}(a) shows $O_{\rm B}$
as a function of $n_{\rm e} (= \phi)$.
The order parameter 
has a pronounced enhancement
in both the high-density region ($0.7 < n_{\rm e} < 1$)
and the low-density one ($0 < n_{\rm e} \; \lsim \; 0.3$).
Figures \ref{orderparameter}(b) and \ref{orderparameter}(c)
depict the $\phi$-dependence of $O_{\rm B}$ for $n_{\rm e}=6/20$
and $16/20$, respectively.
We find a salient growth at $\phi=n_{\rm e}$ for both densities.
We have also observed for $n_{\rm e}=6/20$ that
the order parameter given by Eq.\ (\ref{OB}) with $m=-1$ has a peak
at $\phi=1-n_{\rm e}$.
These results again support the ferromagnetism based on
the statistical transmutation.

Finally, we discuss the region
$0.6 \; \lsim \; n_{\rm e} \; \lsim \; 0.7$,
where the saturated ferromagnetism is absent
in spite of $\phi=n_{\rm e}$.
As a candidate of the competing order, we consider
the spin chirality~\cite{Wen-Wilczek-Zee}
defined by the order parameter
$\chi_{\rm ch} = (1/N) \sum_\ell \langle \mb{S}_\ell \cdot \mb{S}_{\ell + \hat{y}} \times \mb{S}_{\ell + \hat{x}} \rangle$,
where $\hat{x}$ ($\hat{y}$) is the unit vector along $x$ ($y$) direction,
and $\langle \cdots \rangle$ denotes the expectation value in the ground state.
In fact, Nagaoka ferromagnetism in the single-hole case
is known to be destroyed by development of the spin chirality
in the presence of
a perpendicular magnetic field~\cite{Schofield-etal}.
Along the line $\phi = n_{\rm e}$,
we have confirmed that
the spin chirality vanishes when the system exhibits
saturated ferromagnetism.
On the other hand, the chiral order 
is indeed developed
when the saturated ferromagnetism is absent (not shown).

In summary, we have calculated the total spin
of the ground state
in the $U \to \infty$ Hubbard model
with magnetic flux ($\phi$) perpendicular to a square lattice and
revealed regions of saturated ferromagnetism.
The saturated ferromagnetism at $\phi = n_{\rm e}$
is argued to be due to formation of spinful composite bosons.
Statistical transmutation may therefore play a key role in
ferromagnetism in strongly correlated systems, just as it did
in fractional quantum Hall effect.

The present mechanism may be relevant to future experiments on 
an artificial crystal of a square lattice with
quantum dots (i.e., a quantum dot
superlattice)~\cite{Kimura-etal}.
A large lattice constant of such a crystal would enable us
to observe the magnetic-field effect at a modest magnetic field
of a few tesla.
In this situation, the orbital motion rather than the Zeeman effect
could be essential for the emergence of ferromagnetism.
Another possibility is to induce the effective gauge field
internally without an applied magnetic field~\cite{Taguchi-etal}.

We thank D.S. Hirashima and M. Koshino for valuable discussions, and
I. Herbut for critical reading of the manuscript.
Y.S. was supported in part by JSPS Research Fellowships for Young Scientists.
The numerical calculations were performed partly at the Supercomputer Center of the Institute for Solid State Physics, University of Tokyo.
The present work is supported
in part by Grant-in-Aid for Scientific Research and
21st Century COE programs
``Nanometer-Scale Quantum Physics'' at Tokyo Institute of Technology
and ``ORIUM'' at Nagoya University,
all from MEXT of Japan.


\begin{thebibliography}{99}
\bibitem{comment2}
We note that, however, the ``flat-band'' interpretation does not apply
straightforwardly to cases with $\nu <1$, in the light of
the results in Ref.~\cite{Maksym}.
\end{thebibliography}
\end{document}